\documentclass{PoS}
\usepackage{amsmath}

\title{The diagrammatic coaction and the algebraic structure of cut Feynman integrals}

\ShortTitle{The diagrammatic coaction and the algebraic structure of cut Feynman integrals}

\author{\speaker{Samuel Abreu}%
       \\
       Physikalisches Institut, Albert-Ludwigs-Universit\"at
       Freiburg, D-79104 Freiburg, Germany
      E-mail: \email{samuel.abreu@physik.uni-freiburg.de}}

\author{Ruth Britto\\
        School of Mathematics, Trinity College, Dublin 2,
        Ireland ;\\
        Hamilton Mathematics Institute, Trinity
        College, Dublin 2, Ireland ;\\
		Institut de Physique Th\'eorique,
		Universit\'e Paris Saclay,
		CEA, CNRS, F-91191 Gif-sur-Yvette
		cedex, France\\
        E-mail: \email{ruth.britto@tcd.ie}}

\author{Claude Duhr\\
        Theoretical Physics Department, CERN, Geneva,
        Switzerland ;\\
		Center for Cosmology, Particle Physics and
		Phenomenology (CP3),
		Universit\'e Catholique de
		Louvain, 1348 Louvain-La-Neuve, Belgium\\
        E-mail: \email{claude.duhr@cern.ch}}

\author{Einan Gardi\\
        Higgs Centre for Theoretical Physics, School of
        Physics and Astronomy, The University of Edinburgh,
        Edinburgh EH9 3FD, Scotland, UK\\
        E-mail: \email{einan.gardi@ed.ac.uk}}

\abstract{We present a new formula for the coaction of a large
class of integrals. When applied to one-loop (cut) Feynman
integrals, it can be given a diagrammatic representation purely
in terms of pinches and cuts of the edges of the graph. The
coaction encodes the algebraic structure of these integrals, and
offers ways  to extract important properties of complicated integrals
from simpler functions. In particular, it gives direct access to
discontinuities of Feynman integrals and facilitates a
straightforward derivation of the differential equations they
satisfy, which we illustrate in the case of the pentagon.}

\FullConference{13th International Symposium on Radiative
Corrections (Applications of Quantum Field Theory to
Phenomenology)\\
25-29 September, 2017 \\
St.~Gilgen, Austria}

\begin{document}

\section{Introduction}
The realisation that a large class of Feynman integrals can be
written in terms of so-called multiple polylogarithms (MPLs) has
led to major advances in precision calculations in high-energy
physics. A deeper understanding of the algebraic structure of
MPLs has contributed to the development of new efficient techniques
to evaluate Feynman integrals and to handle
the complicated analytic expressions inherent to these
computations, see e.g.~\cite{symbol,diffEq}.
In this context, a very
important tool is the 
coaction \cite{gonchCoaction,brownPeriods}, a mathematical 
operation that exposes properties of MPLs through a 
decomposition into simpler functions. Starting from two loops, 
there are Feynman integrals that cannot be written in terms of 
MPLs only, and extending the coaction to more general classes 
of functions is a pressing question, see e.g.~\cite{elliptic}.
Here, we discuss the
coaction introduced in \cite{diagPRL} and its application to 
one-loop Feynman integrals  where it has a simple diagrammatic 
representation~\cite{diagJHEP}. We illustrate how it
constrains the algebraic structure of these functions by 
discussing its implications in the study of their
discontinuities and the differential 
equations they admit.

\section{A coaction on integrals}

In \cite{diagPRL}, a new formula for the coaction on a large 
class of integrals was proposed which we briefly review here. 
Let $\omega$ be a closed
differential form and $\gamma$ a contour such that the integral 
of $\omega$ over $\gamma$ converges. For simplicity, we also 
assume that $\omega$ vanishes on the boundary of $\gamma$. We 
conjecture that
\begin{equation}\label{eq:masterFormula}
	\Delta\left(\int_\gamma\omega\right)=
	\sum_i\int_\gamma\omega_i\otimes
	\int_{\gamma_i}\omega\,.
\end{equation}
The $\omega_i$ are a basis of master integrands and the
$\gamma_i$ a dual basis of master contours, in the sense that
\begin{equation}\label{eq:ssProj}
	P_{ss}\left(\int_{\gamma_i}\omega_j\right)=\delta_{ij}\,,
\end{equation}
where $P_{ss}$ is a projector onto the subspace of 
`semi-simple' objects $x$ on which the coaction acts trivially, 
$\Delta(x)=x\otimes1$. For the purpose of these proceeding, it 
is sufficient to consider $P_{ss}$ as setting to zero all 
polylogarithmic functions, except where they evaluate to 
$\pi$ or even powers of $\pi$. The second integral on the
right-hand-side
of eq.~(\ref{eq:masterFormula}) is defined mod $i\pi$.

The procedure to construct a coaction based on
eq.~(\ref{eq:masterFormula}) has three main steps:
\begin{enumerate}\setlength\itemsep{0em}
	\item construct a basis of master integrands $\omega_i$ ;
	\item construct a basis of master contours $\Gamma_i$ ;
	\item rotate and normalise the basis $\Gamma_i$ to find a
	new basis $\gamma_i$ dual to $\omega_i$ in the sense of 
	eq.~(\ref{eq:ssProj}).
\end{enumerate}
In the last step we can modify the $\omega_i$ basis 
instead of the $\Gamma_i$, the goal being to construct dual
bases.

The coaction of eq.~(\ref{eq:masterFormula}) interacts with 
discontinuities and differential equations in a very simple 
way. Loosely speaking, the computation of a discontinuity can 
be seen as the modification of the integration contour $\gamma$ 
without modifying the integrand; conversely, differentiation 
modifies the integrand $\omega$ but not the integration 
contour. We thus conclude that discontinuities only act on the 
first entry of the coaction, and derivatives on the second, 
which can be written as:
\begin{equation}\label{eq:discAndDiff}
	\Delta\textrm{Disc}=(\textrm{Disc}\otimes\textrm{id})\Delta
	\qquad\qquad\textrm{and}\qquad\qquad
	\partial\textrm{Disc}=(\textrm{id}\otimes\partial)\Delta\,,
\end{equation}
where id denotes the identity operator.

\subsection{Example: the coaction on multiple polylogarithms}

As a first example of the use of eq.~(\ref{eq:masterFormula}), 
we follow the steps above to reproduce the well-known coaction 
on multiple polylogarithms (MPLs) for generic values of the 
arguments \cite{gonchCoaction,brownPeriods}. MPLs are defined
iteratively as
\begin{equation}
	G(\vec a;z)=\int_0^z\frac{dt}{t-a_1}G(a_2,\ldots;t)
	\equiv\int_0^z\omega_{\vec a}\,,
\end{equation}
and it is easy to see that the set of master integrands 
associated with the integral $G(\vec a;z)$ is $\omega_{\vec b}$ 
with $\vec b\subseteq\vec a$ (we denote the empty subset as
$\vec 0$ and set $\omega_{\vec 0}=dt$).
The obvious candidates for master contours are the paths from 0 
to $z$ that encircle a subset $\vec b$ of
the poles in $\vec a$, which we denote $\Gamma_{\vec b}$,
with $\Gamma_{\vec 0}$ 
the straight line from 0 to $z$. The bases $\omega_{\vec a}$ 
and $\Gamma_{\vec b}$ are not dual to each other in the sense 
of eq.~(\ref{eq:ssProj}), but this can be fixed by normalising 
the contours as $\gamma_{\vec b}=c_{\vec b}\Gamma_{\vec b}$,
with $c_{\vec b}=z^{-1}$ if $\vec b=\vec 0$ and $c_{\vec b}=
(2\pi i)^{-|\vec b|}$ otherwise.
Defining
$G_{\vec b}(\vec a;z)\equiv\int_{\vec b}\omega_{\vec a}$,
we then obtain
\begin{equation}\label{eq:coactionMPLs}
	\Delta G(\vec a;z)=\sum_{\vec b\subseteq \vec a}\int_0^z
	\omega_{\vec b}\otimes
	\int_{\gamma_{\vec b}}\omega_{\vec a} =
	1\otimes G(\vec a;z)+
	\sum_{\vec0\neq\vec b\subseteq \vec a}
	G(\vec b;z)\otimes G_{\vec b}(\vec a;z)\,,
\end{equation}
which reproduces the expected coaction on MPLs
\cite{gonchCoaction,brownPeriods}. We have separated the
term corresponding to $\vec b=\vec 0$ for later reference.

%%%%%%%%%%%%%%%%%%%%%%%%%%%%%%%%%%%%%%%%%%%%%%%%
%%%%%%%%%%%%%%%%%%%%%%%%%%%%%%%%%%%%%%%%%%%%%%%%
%%%%%%%%%%%%%%%%%%%%%%%%%%%%%%%%%%%%%%%%%%%%%%%%

\section{Coaction of one-loop (cut) Feynman integrals}

We conjecture that the coaction in eq.~(\ref{eq:masterFormula}) 
applies to a large class of integrals. Here, we will show how it
applies to one-loop Feynman integrals in dimensional 
regularisation and explore some of the consequences 
\cite{diagPRL,diagJHEP,oneLoopCuts}.
This is a very convenient choice for 
testing the applicability of eq.~(\ref{eq:masterFormula}) 
beyond the case of MPLs. Indeed, since the coefficients of the 
Laurent expansion in the dimensional regulator $\epsilon$ of 
any one-loop integral are conjectured to be MPLs, we can check 
that the prediction of eq.~(\ref{eq:masterFormula}) for the 
coaction on one-loop integrals prior to expansion
reproduces the combinatorics of
the well established coaction on MPLs after expansion.

\paragraph{A basis of master integrands} We consider one-loop 
Feynman integrals with an arbitrary configuration of internal 
and external masses. It is well known that integrals with 
different powers of the propagators satisfy 
integration-by-parts (IBP) relations \cite{ibps},
and integrals in
different spacetime dimensions satisfy dimension-shift 
identities \cite{dimShift}. Tensor integrals can be reduced
through
Passarino-Veltman reduction \cite{pvRed}.
Taking all these relations into 
account, we can choose a basis of master integrands labeled by 
the corresponding Feynman graph $G$ as\footnote{Strictly 
speaking, this is an over complete basis as for sufficiently 
degenerate kinematic configurations there are new IBP 
relations, such as the one relating the one-mass triangle to 
the one-mass bubble. This is not an issue for us because we 
work in dimensional regularisation where these limits are 
smooth.}
\begin{equation}\label{eq:masterIntegrand}
	\omega_{G}=\frac{e^{\gamma_E\epsilon}
	\mathcal{N}_{E_G}}{i\pi^{D_G/2}}d^{D_G}k
	\prod_{j=0}^{|E_G|-1}\frac{1}{(k-q_j)^2-m_j^2}
\end{equation}
where $E_G$ denotes the set of internal edges of $G$ (the 
propagators) and we change the dimensions according to the 
number of propagators,
$D_G=2\lceil|E_G|/2\rceil-2\epsilon$. The $q_j$
are linear combinations of the external momenta $p_i$, and 
the dependence of $\omega_{G}$ on the kinematic invariants
$p_i\cdot p_j$ and $m_j^2$ is implicit. $\mathcal{N}_{E_G}$
is a 
normalisation factor that guarantees that the associated Feynman 
integral has `unit leading singularity' \cite{leadingSing} and 
it can be computed by evaluating the maximal cut at
$\epsilon=0$. In direct analogy with the case of MPLs discussed
above, the 
basis of master integrands relevant to the study of the Feynman 
integral associated with the graph $G$ is given by all
$\omega_{G_C}$, where $C$ is a subset of $E_G$ and $G_C$ is the 
graph obtained by pinching all propagators that are not in $C$.

This choice of master integrands is motivated by the fact that
the corresponding Feynman integrals
\begin{equation}
	J_G=\frac{e^{\gamma_E\epsilon}\mathcal{N}_{E_G}}
	{i\pi^{D_G/2}}\int d^{D_G}k\prod_{j=0}^{|E_G|-1}
	\frac{1}{(k-q_j)^2-m_j^2} \equiv\int_{\Gamma_\emptyset}
	\omega_{G}
\end{equation}
have very nice analytic properties. In particular, they satisfy
differential equations in so-called canonical form \cite{diffEq}
and the
coefficients of their Laurent expansion in $\epsilon$ have
uniform weight.

\paragraph{A basis of master contours} Any two integration
contours are equivalent if one can be smoothly deformed into the
other, without crossing any singularities of the integrand. To
find a basis of master contours, we thus start by studying the
singularities of the integrand, the so-called Landau
singularities \cite{LandauSing}. For one-loop integrals, there
are two kinds of singularities: the singularities of the first
kind, corresponding to a set of propagators being put on-shell, 
and the singularities of the second kind, which also pinch the 
contour at infinity. Candidates for master contours are thus 
the 
contours $\Gamma_{i_1\ldots i_n}$ that encircle the poles of 
propagators $i_1$ through $i_n$, and the contours
$\Gamma_{\infty i_1\ldots i_n}$ which also encircle the pole at 
infinity. The question of whether all these contours are 
independent can be answered in the context of homology theory. 
This was addressed in the 60s \cite{homology} for integer
space-time dimensions, and we have checked that the conclusions still
hold in dimensional regularisation \cite{oneLoopCuts}.
The so-called Decomposition 
Theorem establishes that the contours satisfy the following
relations:
\begin{eqnarray}\label{eq:decomp1}
	&|C|\,\,\textrm{odd:}\qquad\qquad \Gamma_{\infty C}&=
	-2\,\Gamma_C-\sum_{e\in E_G\setminus C}\Gamma_{Ce}+\ldots\,,
	\qquad\qquad \\\label{eq:decomp2}
	&|C|\,\,\textrm{even:}\qquad\qquad \Gamma_{\infty C}&=
	-\sum_{e\in E_G\setminus C}\Gamma_{Ce}-
	\sum_{e,f\in E_G\setminus C}\Gamma_{Cef}+\ldots\,,
\end{eqnarray}
where $C$ is a subset of the set of propagators $E_G$ of a graph
$G$, and we have omitted writing contours that lead to contributions
that vanish mod $i\pi$.
Due to eqs.~\eqref{eq:decomp1} and \eqref{eq:decomp2},
we may exclude the contours that
encircle the
pole at infinity and choose the elements of our basis of 
master contours to be the $\Gamma_{i_1\ldots i_n}$, labeled by 
the subset of propagator poles they encircle.

This choice of master contours is directly related to the 
notion of \emph{cut Feynman integral}. Indeed, we must evaluate
integrals of the form
\begin{equation}\label{resCut}
	\mathcal{C}_CJ_G\equiv\int_{\Gamma_{C}}\omega_G\,,
\end{equation}
which correspond to setting the propagators in $C$ on-shell. In 
the one-loop case, these integrals have been extensively 
discussed in \cite{oneLoopCuts}. We note that the relations in 
eqs.~(\ref{eq:decomp1}) and (\ref{eq:decomp2}) directly 
translate to new relations between cut integrals that do not 
follow from IBP or dimension-shift identities.

\paragraph{A dual basis of master contours} Having identified a
basis of master integrands and master contours, we must now 
make them dual to each other. For $|C|$ even, we have
\begin{equation}
	P_{ss}\left(\int_{\Gamma_{C}}\omega_G\right)=
	\delta_{C,E_G} \qquad
	\Rightarrow \qquad \gamma_C=\Gamma_C\,,
\end{equation}
which is consistent with the fact that the normalisation factor 
$\mathcal{N}_{E_G}$ in eq.~(\ref{eq:masterIntegrand}) is the 
maximal cut evaluated at $\epsilon=0$. For $|C|$ odd, the 
relation with the dual basis is more complicated:
\begin{equation}
	P_{ss}
	\left(\int_{\Gamma_{C}}\omega_G+\frac{1}{2}
	\sum_{e\in E_G\setminus C}\int_{\Gamma_{Ce}}\omega_G\right)
	=\delta_{C,E_G} \qquad
	\Rightarrow \qquad \gamma_C=
	\Gamma_C+\frac{1}{2}\sum_{e\in E_G\setminus C}\Gamma_{Ce}\,.
\end{equation}
The reason for this more complicated relation can be understood
by writing one-loop integrals as integrals over a compact 
quadric in the complex projective space $\mathbb{CP}^{D+1}$. 
Then it becomes clear that members of our basis of master 
integrands taken at $\epsilon=0$
have a simple pole at infinity for $|E_G|$ odd, 
which is absent in the case of $|E_G|$ even. If 
we had allowed ourselves to pick master contours that encircle 
infinity for $|C|$ odd, as would be natural since this extra 
pole is present, then the relation with the dual basis 
would have been simpler. Indeed, using eq.~(\ref{eq:decomp1}) 
we have
\begin{equation}
	\gamma_C=\Gamma_C+\frac{1}{2}\sum_{e\in E_G\setminus C}
	\Gamma_{Ce}
	=-\frac{1}{2}\Gamma_{\infty C}+\ldots
	,\qquad |C|\,\,\textrm{odd}\,,
\end{equation}
where the dots lead to contributions
that vanish mod $i\pi$.

\subsection{Diagrammatic coaction for one-loop
Feynman integrals}

Having constructed dual bases of master integrands and master
contours, we can now use eq.~(\ref{eq:masterFormula}) to write 
the coaction on one-loop Feynman integrals:
\begin{equation}\label{eq:oneLoopMF}
	\Delta(J_G)=\sum_{\emptyset\neq C\subseteq E_G}
	\int_{\Gamma_\emptyset}\omega_{G_C}\otimes
	\int_{\gamma_C}\omega_{G}\,.
\end{equation}
Because master contours and master integrals can be labeled by 
subsets of propagators of the Feynman graph $G$, and given
eq.~\eqref{resCut},
this coaction 
has a very simple diagrammatic formulation:
\begin{equation}\label{eq:diagC}
	\Delta(J_G)=\sum_{\emptyset\neq C\subseteq E_G}J_{G_C}
	\otimes
	\left(\mathcal{C}_CJ_G+a_C\sum_{e\in E_G\setminus C}
	\mathcal{C}_{Ce}J_G\right)\,,
\end{equation}
where $a_C=1/2$ for $|C|$ odd and 0 otherwise. We note that all
our discussion in this section extends trivially to the 
coaction on cut one-loop Feynman integrals, but we will not 
discuss this further in these proceedings.

%%%%%%%%%%%%%%%%%%%%%%%%%%%%%%%%%%%%%%%%%%%%%%%%
%%%%%%%%%%%%%%%%%%%%%%%%%%%%%%%%%%%%%%%%%%%%%%%%
%%%%%%%%%%%%%%%%%%%%%%%%%%%%%%%%%%%%%%%%%%%%%%%%

\section{Checks and applications of the diagrammatic coaction}

The diagrammatic coaction of eq.~(\ref{eq:oneLoopMF}), or 
equivalently eq.~(\ref{eq:diagC}), is based on the conjecture in
eq.~\eqref{eq:masterFormula},
and we must show that it gives consistent results. We now briefly review the several 
checks discussed in \cite{diagJHEP}.

The first nontrivial consistency check of
eq.~(\ref{eq:oneLoopMF}) is to verify that coaction components
of the form $1\otimes J_G$ are correctly reproduced. Indeed, 
the empty set is excluded from the sum in
eq.~(\ref{eq:oneLoopMF}) because $\Gamma_\emptyset$ is not a 
master contour, and we thus only have cut integrals appearing 
in the second entry of the tensor in the coaction.
This is unlike
what happened for MPLs where this term appears explicitly, see
eq.~(\ref{eq:coactionMPLs}). The only way for the coaction to 
be consistent is that there is a linear combination of cut 
integrals that reproduces the uncut integral. This relation can 
be obtained by setting $C=\emptyset$ in eq.~(\ref{eq:decomp2}), 
and then computing the integral over $\Gamma_\infty$. We then 
find the remarkable relation valid for any one-loop integral
\cite{oneLoopCuts}:
\begin{equation}\label{eq:poleCancelationId}
	\sum_{e\in E_G}\mathcal{C}_eJ_G+
	\sum_{\substack{e,f\in E_G\\e<f}}\mathcal{C}_{ef}J_G=
	-\epsilon J_G \mod i \pi\,.
\end{equation}
This relation guarantees that terms of the form 
$1\otimes J_G$ are correctly reproduced by
eq.~(\ref{eq:oneLoopMF}). The coaction component $J_G\otimes1$
can also be checked to be correctly reproduced by
eq.~(\ref{eq:oneLoopMF}) for any one-loop integral.

Aside from consistency checks, we have also verified
eq.~(\ref{eq:oneLoopMF}) in a large number of examples. For 
these, it is important to stress that the coaction is valid in 
dimensional regularisation and prior to expansion in $\epsilon$.
This means
that although we formulated it for a completely generic mass 
configuration, it is still valid in any degenerate kinematic 
configuration: one simply sets to zero scaleless Feynman 
integrals and vanishing cut integrals. As an example, consider 
a three-point function with massless propagators. The pinches 
include scaleless tadpoles which are set to zero, and we thus
obtain from eq.~(\ref{eq:diagC}):
\begin{equation}\label{eq:triangle}
\Delta\left[\!\raisebox{-3.7mm}{
\includegraphics[keepaspectratio=true, width=1.25cm]
{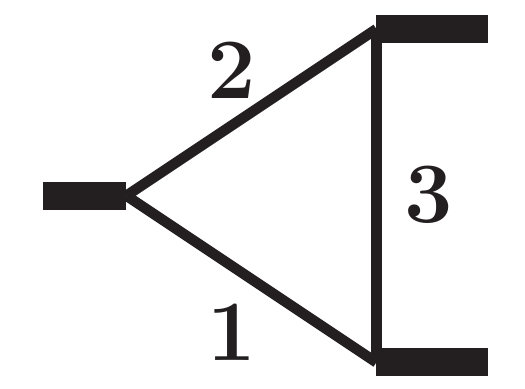}}\right]=\!\!
\raisebox{-2.7mm}
{\includegraphics[keepaspectratio=true, width=1.2cm]
{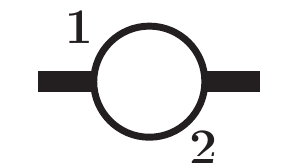}}\!\!\otimes\!\!\raisebox{-3.7mm}
{\includegraphics[keepaspectratio=true, width=1.25cm]
{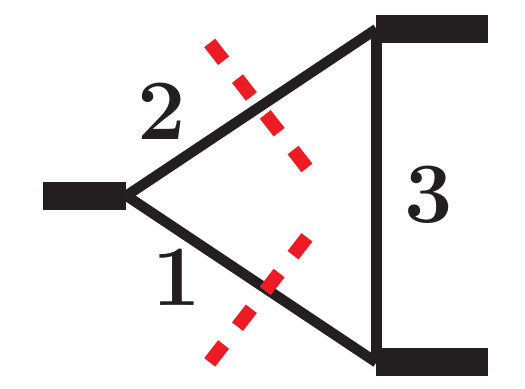}}\!\!
+\!\!\raisebox{-2.7mm}
{\includegraphics[keepaspectratio=true, width=1.2cm]
{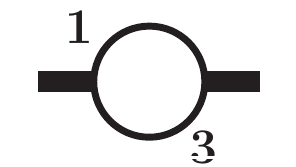}}\!\!\otimes\!\!\raisebox{-3.7mm}
{\includegraphics[keepaspectratio=true, width=1.25cm]
{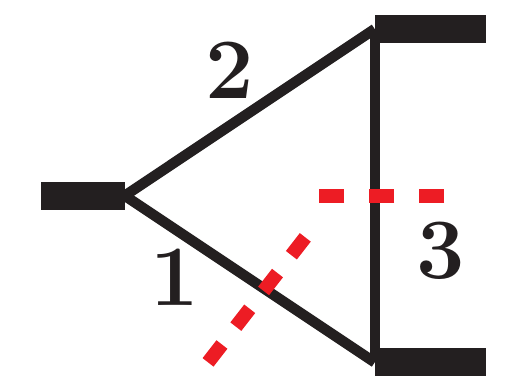}}\!\!
+\!\!\raisebox{-2.7mm}
{\includegraphics[keepaspectratio=true, width=1.2cm]
{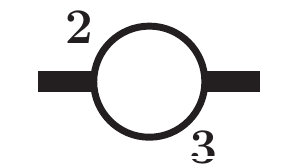}}\!\!\otimes\!\!\raisebox{-3.7mm}
{\includegraphics[keepaspectratio=true, width=1.25cm]
{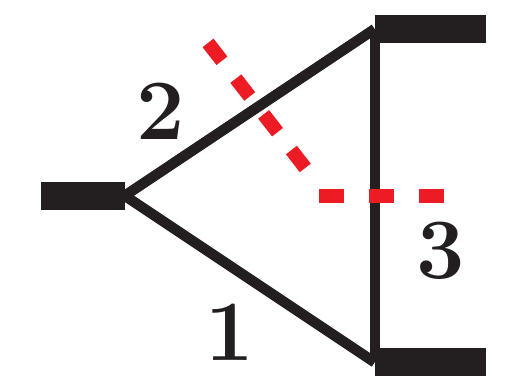}}\!\!
+\!\!\raisebox{-3.7mm}
{\includegraphics[keepaspectratio=true,width=1.25cm]
{./diagrams/triPRL.pdf}}\!\!\otimes\raisebox{-3.7mm}
{\includegraphics[keepaspectratio=true, width=1.25cm]
{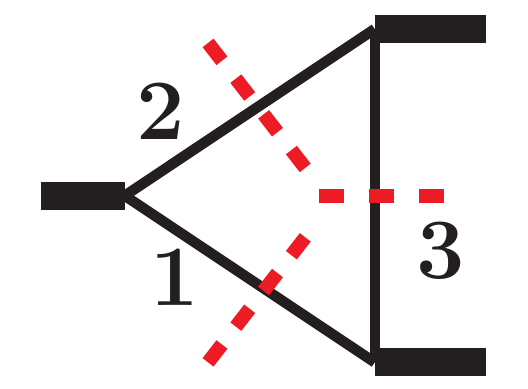}}.
\end{equation}
To verify that \eqref{eq:triangle} is correct,
each integral is expanded in $\epsilon$.
Each term in the expansion is a combination 
of MPLs, of increasing weight as one goes to higher orders in $
\epsilon$. At each order in $\epsilon$, we check that the 
coaction of MPLs acting on both sides of
eq.~(\ref{eq:triangle})
gives the same result. Through this
approach, we reduce the 
check of our conjecture to a check that relies on the coaction 
of MPLs, which was first established from completely different 
arguments on a rigorous mathematics 
footing~\cite{gonchCoaction,brownPeriods}.
For this 
 example, the explicit check was done up to weight 
4. Note that while the left-hand side of eq.~\eqref{eq:triangle}
is finite, the right-hand side has poles associated with 
the bubble integrals. The contribution of these poles cancels 
because of eq.~(\ref{eq:poleCancelationId}). The cancellation
of the poles appearing from bubble or tadpole
integrals due to eq.~\eqref{eq:poleCancelationId} is
another general feature of the diagrammatic coaction.

\subsection{Applications of the diagrammatic coaction}
We now briefly discuss how the diagrammatic coaction encodes 
the algebraic structure of one-loop integrals, and how it can 
be used to obtain important information on these functions in
a simple way. In particular, we would like to highlight the
fact that it constrains the algebraic structure of a
complicated integral through relations to simpler ones,
obtained from its pinches
and cuts.

\subsubsection{Discontinuities}
All known results on discontinuities of Feynman integrals
(see \cite{discs} for some examples) are 
trivially reproduced by the diagrammatic coaction, because the 
first entries of the coaction tensors are Feynman integrals, and
discontinuity operators only act on the first entry of the 
coaction, see eq.~(\ref{eq:discAndDiff}). As an example,
consider the discontinuity on the channel $s_{12}=(q_1-q_2)^2$
of the three point function in eq.~(\ref{eq:triangle}),
with $q_i$ the momentum of propagator $i$ and all $q_i$ moving
clockwise around the diagram.
To compute this discontinuity, one 
simply needs to know how to compute the discontinuity of the 
bubble integral appearing in the first term on the right-hand
side of
eq.~(\ref{eq:triangle}), which is trivial as it is
a single scale 
integral whose discontinuity is just $2\pi i$. We directly 
recover the expected result that the discontinuity function is 
the two-propagator cut (we write $\sim$ because the precise
relation
depends on the definition of Disc):
\begin{equation}
\textrm{Disc}_{s_{12}}\left(\raisebox{-3.7mm}
{\includegraphics[keepaspectratio=true, width=1.25cm]
{./diagrams/triPRL.pdf}}\right)\sim(2\pi i)
\raisebox{-3.7mm}
{\includegraphics[keepaspectratio=true, width=1.25cm]
{./diagrams/triPRLCut12.pdf}}\,.
\end{equation}
The relation between
coaction entries and discontinuities, and in particular the
reason why it is sufficient to consider the term with a bubble
integral in the first entry for
this example, is explained in more detail
in~\cite{diagJHEP}.
The diagrammatic coaction also allows us to present a sharper 
version of the so-called `first-entry-condition': 
In the coaction of a (cut) Feynman integral, the 
first entries are themselves Feynman integrals, with a subset 
of propagators but the same set of cut propagators.

\subsubsection{Differential equations}

Another new insight given by the diagrammatic coaction on the 
algebraic structure of one-loop cut Feynman integrals relates 
to the differential equations they satisfy. The fact that 
differential operators only act on the last entry of the 
coaction, see eq.~(\ref{eq:discAndDiff}), implies that the 
differential equation of any one-loop integral is completely 
determined from its cuts. In fact, it is determined by a very 
small subset of all cuts, and only at very specific orders in
$\epsilon$ \cite{diagJHEP}.
More precisely, the coefficients of the differential 
equation are just the derivatives of the weight-one terms in 
the Laurent expansion of any of the cuts of the integral under 
consideration. For a finite one-loop integral, these are: the 
maximal cut at order $\epsilon^1$, the next-to-maximal cut at 
order $\epsilon^1$ ($\epsilon^0$) for an even (odd) number of 
propagators, and the next-to-next-to-maximal cut at order
$\epsilon^0$. These cuts have been computed for a completely 
generic one-loop integral, and are easily written in terms of 
Cayley and Gram determinants (see eqs.~(9.11) and (9.12) in 
\cite{diagJHEP}). Using the diagrammatic coaction, together with
the explicit results of these cuts, we can thus write down the 
differential equation for any one-loop integral, without 
needing IBP relations as required in more standard approaches. 
For instance, the differential equation for a five-point 
function in our basis, i.e.~in $6-2\epsilon$ dimensions and
normalised to have unit leading singularity, and
with any mass configuration is given by
\begin{align}\begin{split}
\label{eq:pent}
d\!\!\left[\raisebox{-5mm}
{\includegraphics[keepaspectratio=true, width=1.3cm]
{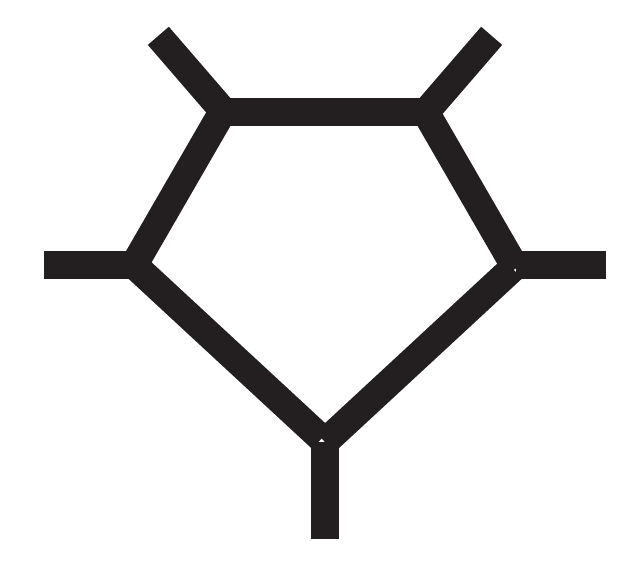}}\right]&=
\sum_{(ijk)}\raisebox{-4mm}
{\includegraphics[keepaspectratio=true, width=1cm]
{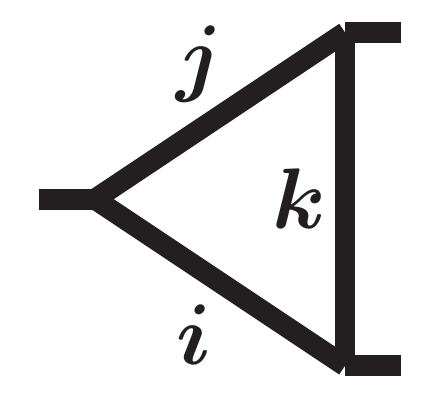}}d\!\!
\left[\raisebox{-5mm}
{\includegraphics[keepaspectratio=true, width=1.3cm]
{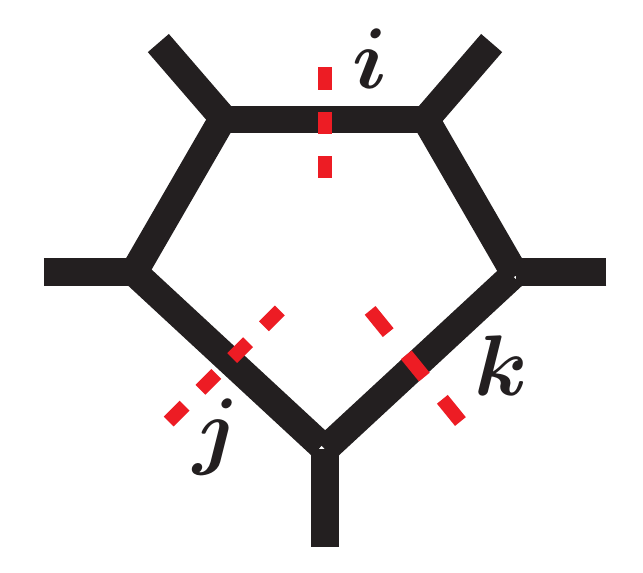}}\!
+\!\frac{1}{2}\sum_{l}\raisebox{-5mm}
{\includegraphics[keepaspectratio=true, width=1.3cm]
{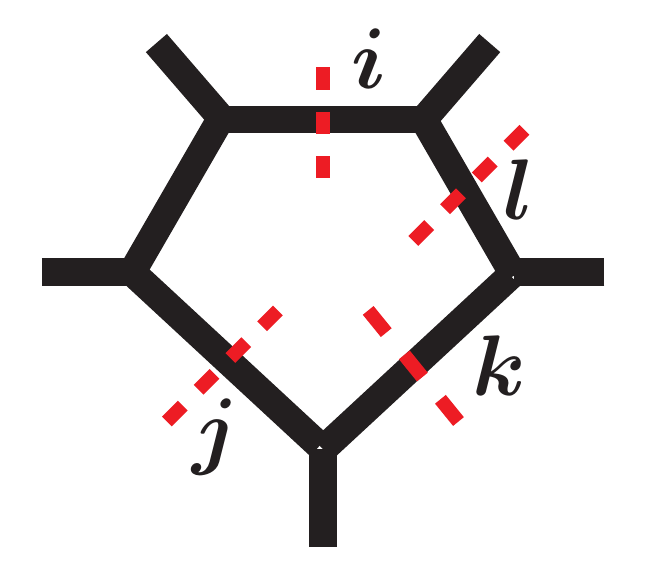}}\right]_{\epsilon^0}\!\!
+\!\sum_{(ijkl)}
\raisebox{-4mm}
{\includegraphics[keepaspectratio=true, width=1.3cm]
{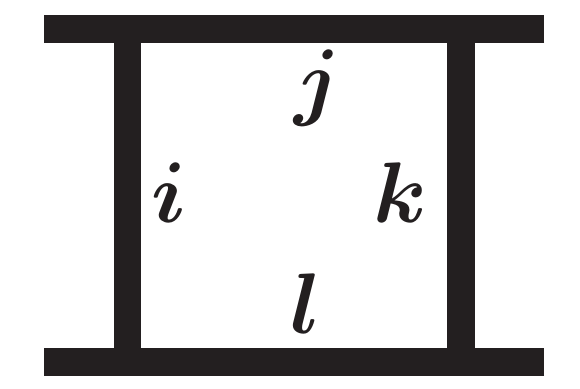}}
d\!\!\left[\raisebox{-5mm}
{\includegraphics[keepaspectratio=true, width=1.3cm]
{./diagrams/pentagonTQuad2.pdf}}\right]_{\epsilon^0}\\
&+\epsilon\,\raisebox{-5mm}
{\includegraphics[keepaspectratio=true, width=1.3cm]
{./diagrams/pentagon2.pdf}}
d\!\!\left[
\raisebox{-5mm}{\includegraphics
[keepaspectratio=true,width=1.3cm]
{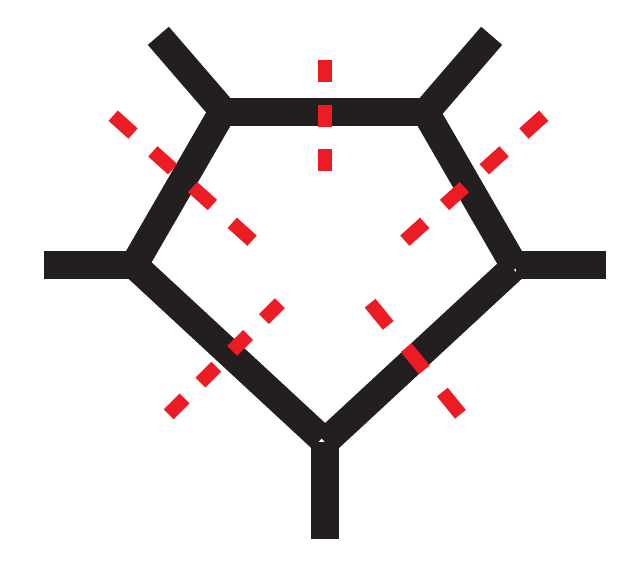}}\right]_{\epsilon^1},
\end{split}\end{align}
where $i$, $j$, $k$ and $l$ run over distinct edges of the 
graph. Since the knowledge of the differential equation 
completely determines the symbol of a function, these cuts
iteratively construct the symbol of any one-loop 
integral (an alternative algorithm was recently proposed in
\cite{Arkani-Hamed:2017ahv}).
This is another example of how the coaction allows us
to gain information on the algebraic structure of complicated 
functions from that of simpler ones.

To make the discussion more concrete, we give the explicit
results for the relevant cuts of the fully massless
pentagon. We will use thin lines to denote massless propagators
and external legs, take all
external momenta as incoming and the
propagator momenta $q_i$, with $i\in\{1,\ldots,5\}$, moving
clockwise around the diagram.
The integral is then a function of the five invariants
$s_i=(q_i-q_{i+2})^2$, with the indices of the $q_i$ understood
cyclically. The order
$\epsilon^1$ term of the maximal cut is 
\begin{equation}\label{eq:maxPent}
	\raisebox{-5mm}{\includegraphics
	[keepaspectratio=true,width=1.3cm]
	{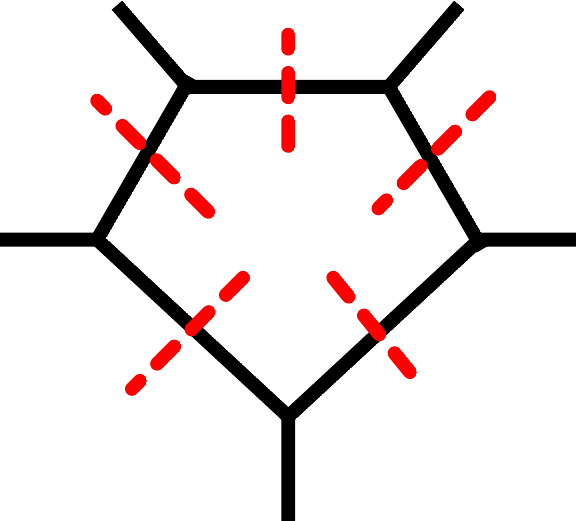}}\,\Bigg\vert_{\epsilon^1}
	=\ln\left(-\frac{16\,\text{Gram}_5}{s_1s_2s_3s_4s_5}\right)
	\,,
\end{equation}
where $\text{Gram}_5=\det(q_i\cdot q_j)$ for $1\leq i,j\leq 4$
is the usual Gram determinant associated with this function.
For the
next-to-maximal cuts, we first define the auxiliary functions
\begin{align}\begin{split}
	&v(x_1,x_2,x_3,x_4,x_5)
	=x_1x_5+x_1x_2+x_3x_4-x_2x_3-x_4x_5
	+\sqrt{16\,\text{Gram}_5}\,,\\
	&\bar{v}(x_1,x_2,x_3,x_4,x_5)
	=x_1x_5+x_1x_2+x_3x_4-x_2x_3-x_4x_5
	-\sqrt{16\,\text{Gram}_5}\,.
\end{split}\end{align}
Then, the order $\epsilon^0$ term of the cut integral where all
propagators but $i$ are cut is given by:
\begin{equation}\label{eq:quadPent}
	\raisebox{-5mm}{\includegraphics
	[keepaspectratio=true,width=1.4cm]
	{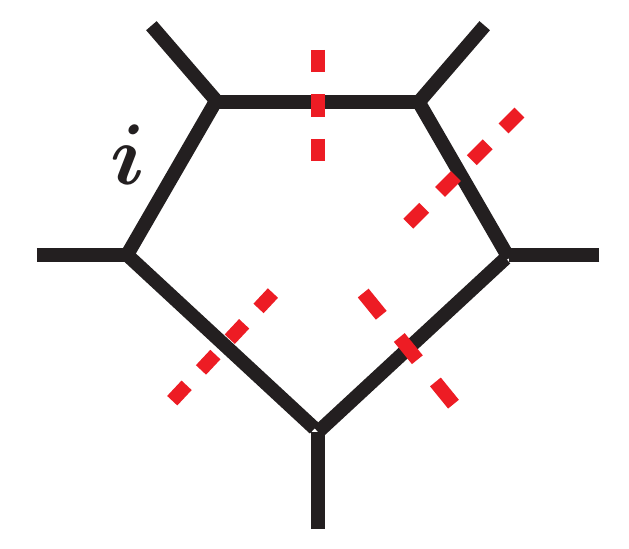}}\,\Bigg\vert_{\epsilon^0}=
	\ln\left(\frac{\bar{v}(s_{i-1},s_i,s_{i+1},s_{i+2},s_{i+3})}
	{v(s_{i-1},s_i,s_{i+1},s_{i+2},s_{i+3})}\right)\,.
\end{equation}
Finally, in the massless case all triple cuts vanish,
\begin{equation}\label{eq:tripPent}
	\raisebox{-5mm}{\includegraphics
	[keepaspectratio=true,width=1.3cm]
	{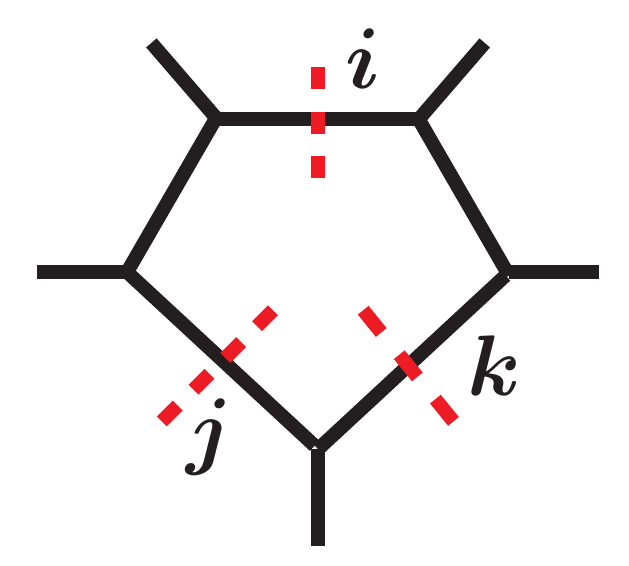}}\,=0\qquad
	\forall\,\,i,j,k\in\{1,2,3,4,5\}.
\end{equation}
Using eqs.~\eqref{eq:maxPent}, \eqref{eq:quadPent}
and \eqref{eq:tripPent} in the generic expression
\eqref{eq:pent}, we obtain the differential equation for the
massless pentagon, valid to all orders in $\epsilon$. Noting
that in the massless case all triangles in \eqref{eq:pent}
are reducible to bubbles, we reproduce the expected differential
equation \cite{diffEqPent}. As another check, we note that the 
letters in eqs.~\eqref{eq:maxPent} and \eqref{eq:quadPent} are a
subset of those appearing in two-loop five-point massless
integrals \cite{letters5Pts}.

%%%%%%%%%%%%%%%%%%%%%%%%%%%%%%%%%%%%%%%%%%%%%%%%
%%%%%%%%%%%%%%%%%%%%%%%%%%%%%%%%%%%%%%%%%%%%%%%%
%%%%%%%%%%%%%%%%%%%%%%%%%%%%%%%%%%%%%%%%%%%%%%%%

\section{Conclusion}

We have presented a new formula for a coaction on a large class 
of integrals. When applied to one-loop Feynman integrals, it 
has a very simple diagrammatic interpretation in terms of cuts 
and pinches of the original integrals. The consistency of the
diagrammatic coaction is checked by showing that it reproduces the 
combinatorics of the coaction on MPLs acting on the 
coefficients of the Laurent expansion of one-loop integrals. 
We then discussed how the diagrammatic coaction encodes 
important information on the algebraic structure of one-loop
integrals. In particular, we argued that
it reproduces known results on the discontinuities of these  
functions and showed how it completely determines the
differential equation
they satisfy. The application of the general formula for the 
coaction to other classes of functions, 
including for Feynman integrals beyond one-loop that do not 
evaluate to MPLs,  is ongoing work.

\begin{acknowledgments}
This work is supported by
the Alexander von Humboldt Foundation
in the framework of the Sofja Kovalevskaja Award 2014, endowed
by the German Federal Ministry of Education and Research (SA),
the ERC Consolidator Grant 647356 ``CutLoops'' (RB), the ERC 
Starting Grant 637019 ``MathAm'' (CD), and the STFC 
Consolidated Grant
``Particle Physics at the Higgs Centre''~(EG).
\end{acknowledgments}

\end{document}